\documentstyle[11pt, amssymb]{article}

\begin{document}

\begin{flushleft}
{\Large {\bf Gluing theory of Riemann surfaces and Liouville conformal field theory}} 
\end{flushleft}

\begin{flushleft}
{\large {\bf Takashi Ichikawa}}  
\end{flushleft}

\noindent 
{\small Department of Mathematics, Graduate School of Science and Engineering, 
Saga University, Saga 840-8502, Japan. 
E-mail: ichikawn@cc.saga-u.ac.jp} 
\vspace{2ex}

\noindent
{\bf Abstract:} 
We study the gluing theory of Riemann surfaces using formal algebraic geometry, 
and give computable relations between the associated parameters 
for different gluing processes. 
As its application to the Liouville conformal field theory, 
we construct the sheaf of tempered conformal blocks 
on the moduli space of pointed Riemann surfaces which satisfies 
the factorization property and has a natural action of the Teichm\"{u}ller groupoid. 
\vspace{2ex}


\begin{flushleft}
{\bf 1. Introduction} 
\end{flushleft} 

\noindent
The Liouville conformal field theory is well studied since it is an important example 
of non-rational conformal field theories, 
and there are remarkable relations pointed by physicists with 
the quantum Teichm\"{u}ller theory (cf. \cite{V, T3, T5}) 
and the $4$ dimensional gauge theory (cf. \cite{AGT}). 
A basic tool in the study of the Liouville theory is to consider expansions of 
local conformal blocks by gluing parameters of Riemann surfaces. 
For example, the AGT correspondence conjectures the coincidence between 
these expansions and instanton partition functions. 
Furthermore, Teschner \cite{T1, T2, T3, T4, T5} claims that 
by studying analytic continuations of these expansions, 
one may obtain spaces of global conformal blocks satisfying the factorization principle. 
The aim of this paper is to apply the gluing theory of Riemann surfaces 
to the study of Liouville conformal blocks, especially of Teschner's consideration. 

First, we study the gluing theory of Riemann surfaces using formal algebraic geometry, 
and give computable relations between the associated parameters 
for different gluing processes. 
By this result, one can study arithmetic geometry of Teichm\"{u}ller groupoids which 
was introduced by Grothendieck \cite{G}, and studied by Moore-Seiberg \cite{MS} and others 
\cite{BK1, BK2, FG, G, HLS, NS}. 
Second, by studying analytic continuations of (local) gluing conformal blocks, 
we construct (generally infinite dimensional) Hilbert spaces consisting of 
``tempered'' Liouville conformal blocks. 
More precisely, 
these Hilbert spaces give a sheaf of conformal blocks, 
namely a vector bundle with projectively flat connection 
on the moduli space of pointed Riemann surfaces which satisfies 
the factorization property and has a natural action of the Teichm\"{u}ller groupoid. 
Therefore, we can give a mathematical foundation of considerations by Teschner 
\cite{T3, T4, T5} on the ``modular functor conjecture'', 
namely that there exists a global theory of Liouville conformal blocks 
which gives a modular functor in the sense of Segal \cite{Se}. 

The organization of this paper is as follows. 
In Section 2, we recall results of \cite{I1, I2} on computable relations between 
deformation parameters of degenerate (algebraic) curves 
which are used in Section 3 to construct Teichm\"{u}ller groupoids 
in the category of arithmetic geometry. 
In Section 4, by combining these results with results of Teschner \cite{T1, T2, T3} and 
Hadasz-Jask\'{o}lski-Suchanek \cite{HJS}, 
we construct the sheaf of tempered Liouville conformal blocks. 
\vspace{2ex}

\begin{flushleft} 
{\bf 2. Deformation of degenerate curves}  
\end{flushleft}

\noindent
{\it 2.1. Degenerate curve.} 
We recall the well known correspondence 
between certain graphs and degenerate pointed curves, 
where a (pointed) curve is called {\it degenerate} if it is a stable (pointed) curve and 
the normalization of its irreducible components are all projective (pointed) lines. 
A {\it graph} $\Delta = (V, E, T)$ means a collection  
of 3 finite sets $V$ of vertices, $E$ of edges, $T$ of tails 
and 2 boundary maps 
$$
b : T \rightarrow V, 
\ \ b : E \longrightarrow \left( V \cup \{ \mbox{unordered pairs of elements of $V$} \} \right) 
$$
such that the geometric realization of $\Delta$ is connected. 
A graph $\Delta$ is called {\it stable} 
if its each vertex has degree $\geq 3$, 
i.e. has at least $3$ branches. 
Then for a degenerate pointed curve, 
its dual graph $\Delta = (V, E, T)$ by the correspondence: 
$$
\begin{array}{lcl}
V & \longleftrightarrow & 
\{ \mbox{irreducible components of the curve} \}, 
\\
E & \longleftrightarrow & 
\{ \mbox{singular points on the curve} \}, 
\\
T & \longleftrightarrow & 
\{ \mbox{marked points on the curve} \} 
\end{array}
$$
such that an edge (resp. a tail) of $\Delta$ has a vertex as its boundary 
if the corresponding singular (resp. marked) point belongs 
to the corresponding component. 
Denote by $|X|$ the number of elements of a finite set $X$. 
Under fixing a bijection 
$\nu : T \stackrel{\sim}{\rightarrow} \{ 1, ... , |T| \}$, 
which we call a numbering of $T$, 
a stable graph $\Delta = (V, E, T)$ becomes the dual graph 
of a degenerate $|T|$-pointed curve of genus 
${\rm rank}_{\mathbb Z} H_{1}(\Delta, {\mathbb Z})$ 
and that each tail $h \in T$ corresponds to the $\nu(h)$th marked point. 
In particular, a stable graph without tail is the dual graph of 
a degenerate (non-pointed) curve by this correspondence. 
If $\Delta$ is trivalent, i.e. any vertex of $\Delta$ has just $3$ branches, 
then a degenerate $|T|$-pointed curve with dual graph $\Delta$ 
is maximally degenerate. 
\vspace{2ex}

\noindent
{\it 2.2. Generalized Tate curve.} 
Let $\Delta = (V, E)$ be a stable graph without tail, 
and under an orientation of $\Delta$, 
i.e., an orientation of each $e \in E$, 
denote by $v_{h}$ the terminal vertex of $h \in \pm E$ 
(resp. the boundary vertex $b(h)$ of $h \in T)$. 
Take a subset ${\cal E}$ of $\pm E = \{ e, -e \ | \ e \in E \}$ 
whose complement ${\cal E}_{\infty}$ satisfies the condition that 
$$
{\cal E}_{\infty} 
\cap 
\{ -h \ | \ h \in {\cal E}_{\infty} \} 
\ = \ 
\emptyset, 
$$ 
and that $v_{h} \neq v_{h'}$ for any distinct $h, h' \in {\cal E}_{\infty}$. 
We attach variables $\alpha_{h}$ for $h \in {\cal E}$ 
and $q_{e} = q_{-e}$ for $e \in E$. 
Let $A_{0}$ be the ${\mathbb Z}$-algebra generated by 
$\alpha_{h}$ $(h \in {\cal E})$, 
$1/(\alpha_{e} - \alpha_{-e})$ $(e, -e \in {\cal E})$ 
and $1/(\alpha_{h} - \alpha_{h'})$ 
$(h, h' \in {\cal E}$ with $h \neq h'$ and $v_{h} = v_{h'})$, 
and let 
$$ 
A \ = \ A_{0} [[q_{e} \ (e \in E)]], \ \ 
B \ = \ A \left[ \prod_{e \in E} q_{e}^{-1} \right]. 
$$ 
According to \cite[Section 2]{I1}, 
we construct the universal Schottky group $\Gamma$ 
associated with oriented $\Delta$ and ${\cal E}$ as follows. 
For $h \in \pm E$, 
let $\phi_{h}$ be the element of $PGL_{2}(B) = GL_{2}(B)/B^{\times}$ given by 
$$
\phi_{h} \ = \ 
\frac{1}{\alpha_{h} - \alpha_{-h}} 
\left( \begin{array}{cc} 
\alpha_{h} - \alpha_{-h} q_{h} & - \alpha_{h} \alpha_{-h} (1 - q_{h}) 
\\ 1 - q_{h} & -\alpha_{-h} + \alpha_{h} q_{h} 
\end{array} \right) 
\ {\rm mod}(B^{\times}), 
$$
where $\alpha_{h}$ (resp. $\alpha_{-h})$ means $\infty$ 
if $h$ (resp. $-h)$ belongs to ${\cal E}_{\infty}$. 
Then 
$$
\frac{\phi_{h}(z) - \alpha_{h}}{z - \alpha_{h}} 
\ = \ 
q_{h} \frac{\phi_{h}(z) - \alpha_{-h}}{z - \alpha_{-h}} 
\ \ (z \in {\mathbb P}^{1}), 
$$
where $PGL_{2}$ acts on ${\mathbb P}^{1}$ by linear fractional transformation. 
 
For any reduced path $\rho = h(1) \cdot h(2) \cdots h(l)$ 
which is the product of oriented edges $h(1), ... ,h(l)$ 
such that $h(i) \neq -h(i+1)$ and $v_{h(i)} = v_{-h(i+1)}$, 
one can associate an element $\rho^{*}$ of $PGL_{2}(B)$ 
having reduced expression 
$\phi_{h(l)} \phi_{h(l-1)} \cdots \phi_{h(1)}$. 
Fix a base point $v_{b}$ of $V$, 
and consider the fundamental group 
$\pi_{1} (\Delta, v_{b})$ which is a free group 
of rank $g = {\rm rank}_{\mathbb Z} H_{1}(\Delta, {\mathbb Z})$. 
Then the correspondence $\rho \mapsto \rho^{*}$ 
gives an injective anti-homomorphism 
$\pi_{1} (\Delta, v_{b}) \rightarrow PGL_{2}(B)$ 
whose image is denoted by $\Gamma$. 
It is shown in \cite[Section 3]{I1} 
(and had been shown in \cite[Section 2]{IN} when $\Delta$ is trivalent and has no loop) 
that for any stable graph $\Delta = (V, E)$ without tail, 
there exists a stable curve $C_{\Delta}$ of genus $g$ over $A$ 
which satisfies the following: 

\begin{itemize}

\item 
The closed fiber $C_{\Delta} \otimes_{A} A_{0}$ of $C_{\Delta}$ 
given by putting $q_{e} = 0$ $(e \in E)$ 
is the degenerate curve over $A_{0}$ with dual graph $\Delta$ which is 
obtained from $P_{v} := {\mathbb P}^{1}_{A_{0}}$ $(v \in V)$ 
by identifying $\alpha_{e} \in P_{v_{e}}$ and $\alpha_{-e} \in P_{v_{-e}}$ ($e \in E$), 
where $\alpha_{h} = \infty$ if $h \in {\cal E}_{\infty}$. 

\item 
$C_{\Delta}$ gives a universal deformation of $C_{\Delta} \otimes_{A} A_{0}$. 

\item 
$C_{\Delta} \otimes_{A} B$ is smooth over $B$ 
and is Mumford uniformized (cf. \cite{M}) by $\Gamma$. 

\item 
Let $\alpha_{h}$ $(h \in {\cal E})$ be complex numbers 
such that $\alpha_{e} \neq \alpha_{-e}$ 
and that $\alpha_{h} \neq \alpha_{h'}$ if $h \neq h'$ and $v_{h} = v_{h'}$. 
Then for sufficiently small complex numbers $q_{e} \neq 0$ $(e \in E)$, 
$C_{\Delta}$ becomes a Riemann surface which is Schottky uniformized (cf. \cite{S}) 
by $\Gamma$. 

\end{itemize}

We apply the above result to construct a uniformized deformation of 
a degenerate pointed curve which had been done by Ihara and Nakamura 
(cf. \cite[Section 2, Theorems 1 and 10]{IN}) when the degenerate pointed curve 
is maximally degenerate and consists of smooth pointed projective lines. 
Let $\Delta = (V, E, T)$ be a stable graph with numbering $\nu$ of $T$. 
We define its extension $\tilde{\Delta} = ( \tilde{V}, \tilde{E} )$ 
as a stable graph without tail by adding a vertex with a loop to the end 
distinct from $v_{h}$ for each tail $h \in T$. 
Then from the uniformized curve associated with $\tilde{\Delta}$, 
by substituting $0$ for the deformation parameters which correspond to 
$e \in \tilde{E} - E$ and by replacing the singular projective lines 
which correspond to $v \in \tilde{V} - V$ with marked points, 
one has the required universal deformation. 
\vspace{2ex}

\noindent
{\it 2.3. Comparison of deformations.}
A {\it rigidification} of an oriented stable graph $\Delta = (V, E, T)$ with numbering
$\nu$ of $T$ means a collection $\tau = \left( \tau_{v} \right)_{v \in V}$ 
of injective maps
$$
\tau_{v} : \{ 0, 1, \infty \} \rightarrow 
\left\{ h \in \pm E \cup T \ | \ v_{h} = v \right\} 
$$
such that $\tau_{v}(a) \neq - \tau_{v'}(a)$ for any $a \in \{ 0, 1, \infty \}$ 
and distinct elements $v, v' \in V$ with $\tau_{v}(a), \tau_{v'}(a) \in \pm E$. 
One can see that any stable graph has a rigidification by the induction
on the number of edges and tails. 
Let $\Delta = (V, E, T)$ be a stable graph with numbering of $T$ 
such that only one vertex, which we denote by $v_{0}$, 
has 4 branches and that the other vertices have $3$ branches. 
Fix an orientation of $\Delta$, 
and denote by $h_{1}$, $h_{2}$, $h_{3}$, $h_{4}$ the mutually different elements of 
$\pm E \cup T$ with terminal vertex $v_{0}$. 
Then one can take a rigidification $\tau = (\tau_{v})_{v \in V}$ of $\Delta$ such that 
$$
\tau_{v_{0}}(0) = h_{2}, \ \ \tau_{v_{0}}(1) = h_{3}, \ \ \tau_{v_{0}}(\infty) = h_{4}, 
$$
and hence $x = x_{h_{1}}$ gives the coordinate on 
${\mathbb P}_{\mathbb Z}^{1} - \{ 0, 1, \infty \}$. 
Denote by $C_{(\Delta, \tau)}$ the uniformized deformation given in 2.2 
which is a stable $|T|$-pointed curve over 
$$
A_{(\Delta, \tau)} = 
{\mathbb Z} \left[ x,  x^{-1}, (1-x)^{-1} \right] [[ y_{e} (e \in E) ]]. 
$$
Let $\Delta' = (V', E', T')$ (resp. $\Delta'' = (V'', E'', T'')$) be the trivalent graphs 
obtained from $\Delta = (V, E, T)$ by replacing $v_{0}$ 
with an edge $e'_{0}$ (resp. $e''_{0}$) having two boundary vertices 
one of which is a boundary of $h_{1}$, $h_{2}$ (resp. $h_{1}$, $h_{3}$) 
and another is a boundary of $h_{3}$, $h_{4}$ (resp. $h_{2}$, $h_{4}$). 
Then one can identify $T'$, $T''$ with $T$ naturally, 
and it is easy to see that according as $x \rightarrow 0$ (resp.  $x \rightarrow 1$), 
the degenerate $|T|$-pointed curve corresponding to $x$ becomes 
the maximally degenerate $|T|$-pointed curve 
with dual graph $\Delta'$ (resp. $\Delta''$). 
Let $\Delta'$ (resp. $\Delta''$) without $e'_{0}$ (resp. $e''_{0}$) have the orientation  naturally induced from that of $\Delta$, 
and let $h'_{0}$ (resp. $h''_{0}$) be the edge $e'_{0}$ (resp. $e''_{0}$) with orientation. 
For $i = 1, 2, 3, 4$, we denote by $h'_{i}$ (resp. $h''_{i}$) the oriented edge in 
$\Delta'$ (resp. $\Delta''$) corresponding to $h_{i}$, 
and identify the invariant part
$$
E^{\rm inv} = E - \left\{ |h_{i}| \ | \ 1 \leq i \leq 4 \right\} 
$$	
of $E$ as that of $E'$ and $E''$. 
Then as seen above, 
for a rigidification $\tau'$ (resp. $\tau''$) of $\Delta'$ (resp. $\Delta''$), 
we have the uniformized deformation $C_{(\Delta', \tau')}$ (resp. $C_{(\Delta'', \tau'')}$) 
which is a stable $|T|$-pointed curve over 
$$
A_{(\Delta', \tau')} = {\mathbb Z} [[ s_{e'} (e' \in E') ]] \ \ 
\mbox{(resp. $A_{(\Delta'', \tau'')} = {\mathbb Z} [[ t_{e''} (e'' \in E'') ]]$)}. 
$$
Then we will consider two isomorphisms of $C_{(\Delta, \tau)}$ to $C_{(\Delta', \tau')}$ 
and to $C_{(\Delta'', \tau'')}$. 
Note that under the isomorphisms, 
the comparison between parameters of the base rings depends on the situation 
whether some $h_{i}$ $(1 \leq i \leq 4)$ are loops or not. 
In Theorem 2.1 below, 
we make the comparison in restricted cases for the saving of space 
since the other cases are seen to be treated similarly from the proof. 
\vspace{2ex}

\noindent
{\bf Theorem 2.1.} (cf. \cite[Theorem 1]{I2})  
\begin{it} 
Put $I = \{ 1 \leq i \leq 4 \ | \ h_{i} \in \pm E \}$, 
denote by $y_{i}$ the deformation parameters associated with $h_{i}$ for $i \in I$, 
and denote by $s_{j}$ (resp. $t_{j}$) the deformation parameters 
associated with $h'_{j}$ (resp. $h''_{j}$) for $j \in \{ 0 \} \cup I$. 
Then we have 
\begin{itemize}

\item[{\rm (1)}] 
Over ${\mathbb Z}((x)) [[y_{e} (e \in E)]]$, 
$C_{(\Delta, \tau)}$ is isomorphic to $C_{(\Delta', \tau')}$, 
where under this isomorphism,   
the variables of the base rings $A_{(\Delta, \tau)}$ and $A_{(\Delta', \tau')}$ 
are related as
$$
\frac{x}{s_{0}}, \ \ \frac{y_{i}}{s_{0} s_{i}} \ (i \in \{ 1, 2 \} \cap I), \ \ 
\frac{y_{i}}{s_{i}} \ (i \in \{ 3, 4 \} \cap I), \ \ \frac{y_{e}}{s_{e}} \ (e \in E^{\rm inv}) 
$$
belong to $\left( A_{(\Delta', \tau')} \right)^{\times}$ 
if $|h_{i}|$ $(1 \leq i \leq 4)$ are mutually different, 
and 
$$
\frac{x}{s_{0}}, \ \ \frac{y_{i}}{s_{i}} \ (i \in I), \ \ \frac{y_{e}}{s_{e}} \ (e \in E^{\rm inv}) 
$$
belong to $\left( A_{(\Delta', \tau')} \right)^{\times}$ if $|h_{1}| = |h_{2}|$. 

\item[{\rm (2)}] 
Over ${\mathbb Z}((1-x)) [[y_{e} (e \in E)]]$, 
$C_{(\Delta, \tau)}$ is isomorphic to $C_{(\Delta'', \tau'')}$, 
where under this isomorphism,   
the variables of the base rings $A_{(\Delta, \tau)}$ and $A_{(\Delta'', \tau'')}$ 
are related as
$$
\frac{1-x}{t_{0}}, \ \ \frac{y_{i}}{t_{0} t_{i}} \ (i \in \{ 1, 3 \} \cap I), \ \ 
\frac{y_{i}}{t_{i}} \ (i \in \{ 2, 4 \} \cap I), \ \ \frac{y_{e}}{t_{e}} \ (e \in E^{\rm inv}) 
$$
belong to $\left( A_{(\Delta'', \tau'')} \right)^{\times}$ 
if $|h_{i}|$ $(1 \leq i \leq 4)$ are mutually different, 
$$
\frac{1-x}{t_{0}}, \ \ \frac{y_{i}}{t_{0} t_{i}} \ (i \in \{ 1, 2, 3 \} \cap I), \ \ 
\frac{y_{i}}{t_{i}} \ (i \in \{ 4 \} \cap I), \ \ \frac{y_{e}}{t_{e}} \ (e \in E^{\rm inv}) 
$$
belong to $\left( A_{(\Delta'', \tau'')} \right)^{\times}$ 
if $|h_{1}| = |h_{2}|$, $|h_{3}| \neq |h_{4}|$, 
and 
$$
\frac{1-x}{t_{0}}, \ \ \frac{y_{i}}{t_{0} t_{i}} \ (i \in I), 
\ \ \frac{y_{e}}{t_{e}} \ (e \in E^{\rm inv}) 
$$
belong to $\left( A_{(\Delta'', \tau'')} \right)^{\times}$ 
if $|h_{1}| = |h_{2}|$, $|h_{3}| = |h_{4}|$. 

\end{itemize}
\end{it}

\noindent
{\it Remark 1.} 
In (1) and (2) above, 
the constant terms of the ratios in $A_{(\Delta', \tau')}$, $A_{(\Delta'', \tau'')}$ are 
clearly either $1$ or $-1$, 
and these signs can be easily determined from the data of rigidifications. 
If $|h_{1}| = |h_{2}|$ in (2) particularly, 
then $y_{1}/(t_{0} t_{1})$ belongs to $\left\{ \left( A_{(\Delta'', \tau'')} \right) \right\}^{2}$  
and hence this constant term is $1$ since by \cite[Proposition 1.3]{I1}, 
the reduced element $\phi_{h(l)} \phi_{h(l-1)} \cdots \phi_{h(1)}$ has the multiplier in 
$\prod_{i = 1}^{l} y_{h(i)} \cdot \left( A^{\times} \right)^{2}$ if $h(1) \neq h(l)$. 
\vspace{2ex}

\noindent
{\it Remark 2.} 
From the properties of generalized Tate curves given in 2.2, 
one can see that the assertion in (1) (resp. (2)) holds 
in the category of complex geometry when $x$, $y_{e}$ and $s_{e'}$ 
(resp. $1-x$, $y_{e}$ and $t_{e''}$) are sufficiently small. 
\vspace{2ex}

\noindent
{\it Proof.} 
We review the proof given in \cite{I2} since it also gives the method of 
comparing deformation parameters of degenerate curves. 
We prove the theorem when $\Delta$ has no tail from which 
the assertion in general case follows, 
and we only prove (1) since (2) can be shown in the same way. 
Over a certain open subset of
$$
\left\{ \left( x, y_{e} (e \in E) \right) \ | \ 
x \in {\mathbb C}^{\times}, \ y \in {\mathbb C} \right\} 
$$
with sufficiently small absolute values $|x|$ and $|y_{e}|$, 
$C_{(\Delta, \tau)}$ gives a deformation of the degenerate curve 
with dual graph $\Delta$. 
Hence there exists an isomorphism
$$
{\mathbb C}((x)) [[y_{e} (e \in E)]] \cong {\mathbb C}((s_{0})) [[s_{e'} (e' \neq e'_{0})]]  
$$
which induces an isomorphism $C_{(\Delta, \tau)} \cong C_{(\Delta', \tau')}$ 
such that the degenerations of $C_{(\Delta, \tau)}$ 
given by $y_{i} \rightarrow 0$ $(1 \leq i \leq 4)$ 
and $y_{e} \rightarrow 0$ $(e \in E^{\rm inv})$ 
correspond to those of  of $C_{(\Delta', \tau')}$ 
given by $s_{i} \rightarrow 0$ and $s_{e} \rightarrow 0$ respectively. 
Since these two curves are Mumford uniformized, 
a result of Mumford \cite[Corollary 4.11]{M} implies that 
the uniformizing groups $\Gamma_{(\Delta, \tau)}$ and $\Gamma_{(\Delta', \tau')}$ 
of $C_{(\Delta, \tau)}$ and $C_{(\Delta', \tau')}$ respectively 
are conjugate over the quotient field of
${\mathbb C}((x)) [[y_{e} (e \in E)]] \cong {\mathbb C}((s_{0})) [[s_{e'} (e' \neq e'_{0})]]$. 
Denote by 
$\iota : \Gamma_{(\Delta, \tau)} \stackrel{\sim}{\rightarrow} \Gamma_{(\Delta', \tau')}$ 
the isomorphism defined by this conjugation. 
Since eigenvalues are invariant under conjugation and the cross ratio
$$
[a; b; c; d] = \frac{(a - c)(b - d)}{(a - d)(b - c)} 
$$
of $4$ points $a$, $b$, $c$, $d$ is invariant under linear fractional transformation, 
one can see the following: 
\begin{itemize}

\item[(A)] 
For any $\gamma \in \Gamma_{(\Delta, \tau)}$, 
the multiplier of $\gamma$ is equal to that of $\iota(\gamma)$ 
via the above isomorphism. 

\item[(B)] 
For any $\gamma_{i} \in \Gamma_{(\Delta, \tau)}$ $(1 \leq i \leq 4)$, 
the cross ratio $\left[ a_{1}, a_{2}; a_{3}, a_{4} \right]$ of the attractive fixed points $a_{i}$  of $\gamma_{i}$ is equal to that of $\iota(\gamma_{i})$ via the above isomorphism. 

\end{itemize}

We consider the case that $|h_{i}|$ $(1 \leq i \leq 4)$ are mutually different. 
$$
A_{1} = {\mathbb Z} \left[ \left[ x, \ \frac{y_{1}}{x}, \ \frac{y_{2}}{x}, \ y_{3}, \ y_{4}, \ 
y_{e} \ (e \in E^{\rm inv}) \right] \right], 
$$
whose quotient field is denoted by $\Omega_{1}$, 
and let $I_{1}$ be the ideal of $A_{1}$ generated by 
$x$, $y_{1}/x$, $y_{2}/x$, $y_{3}$, $y_{4}$ and $y_{e}$ $(e \in E^{\rm inv})$. 
Then from (A) and (B) as above and results in [I, \S1], 
we will show that the isomorphism descends to $A_{1} \cong A_{(\Delta', \tau')}$, 
where the variables are related as in the statement of Theorem 1 (1). 
We take local coordinates $\xi_{h}$ as 
$$
\xi_{h}(z) = 
\left\{ \begin{array}{ll} 
z - x & \mbox{(if $h = h_{1}$),} 
\\ 
z      & \mbox{(if $h = \tau_{v}(0)$ for some $v \in V$),} 
\\ 
z - 1 & \mbox{(if $h = \tau_{v}(1)$ for some $v \in V$),}  
\\ 
1/z   & \mbox{(if $h = \tau_{v}(\infty)$ for some $v \in V$).}  
\end{array} \right. 
$$
For $z \in {\mathbb P}^{1}(\Omega_{1})$ with $\xi_{h}(z) \in I_{1}$ and 
$h' \in \pm E - \{ h \}$ with $v_{-h'} = v_{h}$, 
by the calculation in the proof of \cite[Lemma 1.2]{I1}, 
one can see that $\xi_{h'} \left( \phi_{h'}(z) \right) \in I_{1}$. 
Hence for $\gamma \in \Gamma_{(\Delta, \tau)}$ with reduced expression 
$\phi_{h(1)} \cdots \phi_{h(l)}$, 
if we take $h \in \pm E - \{ -h(l) \}$ with $v_{h} = v_{-h(l)}$ 
and $z \in {\mathbb P}^{1}(\Omega_{1})$ with $\xi_{h}(z) \in I_{1}$, 
then by \cite[Lemma 1.2]{I1}, 
the attractive fixed point $a$ of $\gamma$ is given by 
$\lim_{n \rightarrow \infty} \gamma^{n}(z)$, and hence $\xi_{h(1)}(a) \in I_{1}$. 
For each $v \in V$, 
fix a path $\rho_{v}$ in $\Delta$ from the base point $v_{b}$ to $v$. 
If $\rho_{i} \in \pi_{1}(\Delta, v_{0})$ has reduced expression $\cdots h_{i}$ 
$(1 \leq i \leq 4)$, 
then the attractive fixed points $a_{i}$ of 
$\gamma_{i} = \left( \rho_{v_{0}}^{*} \right)^{-1} \cdot \rho_{i}^{*} \cdot \rho_{v_{0}}^{*}$ 
satisfy that $[a_{1}, a_{3}; a_{2}, a_{4}] \in x \cdot (A_{1})^{\times}$. 
Furthermore, by Proposition 1.4 and Theorem 1.5 of \cite{I1}, 
the attractive fixed points $a'_{i}$ of $\iota(\gamma_{i})$ satisfy that 
$[a'_{1}, a'_{3}; a'_{2}, a'_{4}] \in s_{0} \cdot \left( A_{(\Delta', \tau')} \right)^{\times}$, 
and hence from (B), 
we have the comparison of $y_{1}/x$ and $s_{1}$ follows from applying (B) to 
$\gamma_{i} = \left( \rho_{v_{0}}^{*} \right)^{-1} \cdot \rho_{i}^{*} \cdot \rho_{v_{0}}^{*}$ 
$(1 \leq i \leq 4)$, 
where $\rho_{i} \in \pi_{1}(\Delta, v_{0})$ has reduced expression 
$$
\left\{ \begin{array}{lcl} 
\rho_{1} & = & \cdots h_{2}, 
\\ 
\rho_{2} & = & \cdots h_{3}, 
\\ 
\rho_{3} & = & \cdots h_{5} \cdot h_{1}, 
\\ 
\rho_{4} & = & \cdots h_{6} \cdot h_{1},  
\end{array} \right. 
$$
for distinct oriented edges $h_{5}$, $h_{6}$ with terminal vertex $v_{-h_{1}}$. 
Similarly, we have the comparison of $y_{2}/x$ 
(resp. $y_{3}$, $y_{4}$, $y_{e}$ $(e \in E^{\rm inv} - \{ \mbox{loops} \})$) and 
$s_{2}$ (resp. $s_{3}$, $s_{4}$, $s_{e}$, 
and further
if $e \in E^{\rm inv}$ is a loop, 
then the comparison of $y_{e}$ and $s_{e}$ follows from applying (A) to
$\gamma = \left( \rho_{v_{e}}^{*} \right)^{-1} \cdot \phi_{e} \cdot \rho_{v_{e}}^{*}$. 
Therefore, the above isomorphism descends to $A_{1} \cong A_{(\Delta', \tau')}$ 
such that under this isomorphism  
$$
\frac{x}{s_{0}}, \ \ \frac{y_{i}}{x s_{i}} \ (i \in \{ 1, 2 \}), \ \ 
\frac{y_{i}}{s_{i}} \ (i \in \{ 3, 4 \}), \ \ \frac{y_{e}}{s_{e}} \ (e \in E^{\rm inv}) 
$$
belong to $\left( A_{(\Delta', \tau')} \right)^{\times}$. 

One can show the assertion in the case that $|h_{1}| = |h_{2}|$ similarly. 
\ $\square$
\vspace{2ex}

The following result was given substantially in \cite[3.1]{I2}, \cite[1.2]{I3}, 
and is explicitly shown here since this is crucial to prove results in Section 4. 
\vspace{2ex}

\noindent
{\bf Theorem 2.2.} 
\begin{it}
Let the notation be as above. 
Then there are elements $u_{i}$ $(1 \leq i \leq 3g - 4)$ of 
$A_{(\Delta, \tau)} = {\mathbb Z} \left[ x, x^{-1}, (1-x)^{-1} \right][[y_{e} (e \in E)]]$ 
such that under $x \rightarrow 0$ (resp. $1$), 
$\{ x, u_{i} \}$ (resp. $\{ 1 - x, u_{i} \}$) give deformation parameters of the closed fibers 
$C'_{0}$ (resp. $C''_{0}$) of $C_{(\Delta', \tau')}$ (resp. $C_{(\Delta'', \tau'')}$), 
namely one has 
$$
A_{(\Delta', \tau')} \cong {\mathbb Z}[[ x, u_{i} (1 \leq i \leq 3g - 4) ]], \ \ 
A_{(\Delta'', \tau'')} \cong {\mathbb Z}[[ 1-x, u_{i} (1 \leq i \leq 3g - 4) ]]. 
$$
\end{it}
\vspace{-2ex}

\noindent
{\it Proof.} 
Assume that all branches starting from $v_{0}$ are not loops. 
We put $y_{i} = y_{|h_{i}|}$ as in Theorem 2.1. 
Take 
$$
u_{i} = \left\{ \begin{array}{lcll} 
y_{1}/x(1 - x) & \mbox{or} & - y_{1}/x(1 - x) & (i = 1), 
\\ 
y_{2}/x         & \mbox{or} & - y_{2}/x          & (i = 2), 
\\
y_{3}/(1 - x)  & \mbox{or} & - y_{3}/(1 - x)   & (i = 3), 
\\ 
y_{4}             & \mbox{or} & - y_{4}             & (i = 4). 
\end{array} \right. 
$$
and $u_{i}$ $(i \geq 5)$ by specifying one of $y_{e}$ and $-y_{e}$ for each $e \in E^{\rm inv}$. 
Then by Theorem 2.1, under $x \rightarrow 0$ (resp. $1$), 
$x$ (resp. $1 - x$) and $u_{i}$ $(1 \leq i \leq 3g - 4)$ are 
deformation parameters of the maximally degenerate pointed curve $C'_{0}$ (resp. $C''_{0}$). 
In the case that there are loops with boundary vertex $v_{0}$, 
one can also take required deformation parameters using Theorem 2.1. 
\ $\square$ 
\vspace{2ex}

\noindent
{\it 2.4. Ihara-Nakamura's deformation.} 
We consider deformations of maximally degenerate curves using 
``standard'' local coordinates on ${\mathbb P}^{1}$ 
which is studied by Ihara and Nakamura \cite{IN} with application to 
Galois theory on arithmetic fundamental groups of curves. 
Let $\Delta = (V, E, T)$ be a trivalent graph such that 
${\rm rank}_{\mathbb Z} H_{1}(\Delta, {\mathbb Z}) = g$, $|T| = n$, 
and $C$ denote the associated degenerate curve over ${\mathbb Z}$ 
which is a union of $P_{v} = {\mathbb P}^{1}$ $(v \in V)$ 
with marked points $\alpha_{t}$ $(t \in T)$, where $v_{t} = v$,  
by identifying $\alpha_{h} \in P_{v_{h}}$ and $\alpha_{-h} \in P_{v_{-h}}$ $(h \in \pm E)$. 
We take a local coordinate $z_{h}$ on each $P_{v_{h}}$ such that 
$$
\left\{ \mbox{marked points and singular points on $P_{v_{h}}$} \right\} = \{ 0, 1, \infty \}, 
$$
and that $z_{h}(\alpha_{h}) = 0$. 
Then one can define the Ihara-Nakamura deformation $C_{\rm IN}$ of $C_{0}$ 
over the ring ${\mathbb Z}[[q_{e}]]$ 
of integral formal power series of variables $q_{e}$ $(e \in E)$ 
by the relation $z_{h} z_{-h} = q_{|h|}$ $(h \in \pm E)$. 
As is shown in \cite[2.4.2]{IN}, for any $v_{0} \in V$, 
for each $\gamma \in \pi_{1}(\Delta, v_{0})$, 
one can associates an element $\gamma^{*} \in PGL_{2}({\mathbb Z}[[q_{e}]])$ 
as follows. 
Let 
$$
v_{0}, \ h_{0}, \ v_{1}, \ h_{1}, ..., v_{d} = v_{0} \ (d \geq 0)
$$
be the reduced path on $\Delta$ representing $\gamma$ 
such that $v_{-h_{i}} = v_{i}$ and $v_{h_{i}} = v_{i+1}$. 
Let $g_{k}$ $(1 \leq k \leq d-1)$ be the element of $PGL_{2}({\mathbb Z})$ 
defined by $z_{h_{k}} = g_{k} \left( z_{h_{k-1}} \right)$, 
and $h_{k}$ $(1 \leq k \leq d-1)$ be the transform $z \mapsto q_{|h_{k}|} z^{-1}$ 
which comes from the relation $z_{h_{k}} = q_{|h_{k}|} z_{-h_{k}}^{-1}$. 
Put 
$$
\gamma^{*} = g_{d} \circ h_{d-1} \circ \cdots \circ g_{1} \circ h_{0} \circ g_{0}, 
$$
where $g_{0}$, $g_{d}$ are defined by $z_{-h_{0}} = g_{0}(z)$, $z = g_{d} \left( z_{h_{d-1}} \right)$. 
Then $\gamma \mapsto \gamma^{*}$ gives a representation 
$\pi_{1}(\Delta, v_{0}) \rightarrow PGL_{2}({\mathbb Z}[[q_{e}]])$
whose image is a Schottky group over ${\mathbb Z}[[q_{e}]]$ as in 2.2. 
Then by the same way as in the proof of Theorem 2.1 especially 
using the assertions (A) and (B), 
one can compare the deformation parameters $q_{e}$ and those given in Theorem 2.1. 
\vspace{2ex}

\begin{flushleft} 
{\bf 3. Teichm\"{u}ller groupoid}  
\end{flushleft}

\noindent
{\it 3.1. Moduli space of curves.} 
We review fundamental facts on the moduli space of pointed curves and 
its compactification \cite{DM, KM, K}. 
Let $g$ and $n$ be non-negative integers such that $n$ and $2g - 2 + n$ are positive. 
Let ${\cal M}_{g, n}$ (resp. ${\cal M}_{g, \vec{n}}$) denote the moduli stacks 
over ${\mathbb Z}$ of proper smooth curves of genus $g$ with $n$ marked points 
(resp. with $n$ marked points having non-zero tangent vectors). 
Then ${\cal M}_{g, \vec{n}}$ becomes a principal $({\mathbb G}_{m})^{n}$-bundle 
on ${\cal M}_{g, n}$. 
Furthermore, let $\overline{\cal M}_{g, n}$ denote the Deligne-Mumford-Knudsen 
compactification of ${\cal M}_{g, n}$ which is defined as the moduli stack 
over ${\mathbb Z}$ of stable curves of genus $g$ with $n$ marked points, 
and $\overline{\cal M}_{g, \vec{n}}$ denote the $({\mathbb A}^{1})^{n}$-bundle 
on $\overline{\cal M}_{g, n}$ containing ${\cal M}_{g, \vec{n}}$ naturally. 
For these moduli stacks ${\cal M}_{*,*}$ and $\overline{\cal M}_{*,*}$, 
${\cal M}_{*,*}^{\rm an}$ and $\overline{\cal M}_{*,*}^{\rm an}$ denote 
the associated complex orbifolds. 
A {\it point at infinity} on ${\cal M}_{g, n}$ (resp. ${\cal M}_{g, \vec{n}}$) is a point on 
$\overline{\cal M}_{g, n}$ (resp. $\overline{\cal M}_{g, \vec{n}}$) which corresponds to 
a maximally degenerate $n$-pointed curve, 
and a {\it tangential point at infinity} is a point at infinity with tangential structure 
over ${\mathbb Z}$. 

We describe the boundary of ${\cal M}_{g, n}$. 
Denote by ${\cal D}_{0}$ the divisor of $\overline{\cal M}_{g,n}$ corresponding to 
singular stable marked curves which are desingularized to 
stable curves of genus $g-1$ with $n+2$ marked points. 
For an integer $i$ with $1 \leq i \leq [g/2]$, 
and $S$ be a subset of $P = \{ 1,..., n \}$ such that 
$2i - 2 + |S|, 2(g-i) - 2 + n - |S|$ are positive. 
denote by ${\cal D}_{i,S}$ the divisor of $\overline{\cal M}_{g,n}$ 
corresponding to singular stable marked curves which are desingularized to 
the sum of pairs of stable curves of genus $i$ with $|S|$ marked points and 
of genus $g-i$ with $n-|S|$ marked points. 
Then $\overline{\cal M}_{g, n} - {\cal M}_{g, n}$ consists of normal crossing divisors 
${\cal D}_{0}, {\cal D}_{i,S}$, 
and hence $\overline{\cal M}_{g, \vec{n}} - {\cal M}_{g, \vec{n}}$ consists of 
the pullbacks of ${\cal D}_{0}, {\cal D}_{i,S}$ by the natural projection 
$\overline{\cal M}_{g, \vec{n}} \rightarrow \overline{\cal M}_{g, n}$ 
which we denote by the same notation. 
\vspace{2ex}

\noindent
{\it 3.2. Teichm\"{u}ller groupoid.} 
The {\it Teichm\"{u}ller groupoid} for ${\cal M}_{g, \vec{n}}$ is defined as 
the fundamental groupoid for ${\cal M}_{g, \vec{n}}^{\rm an}$ 
with tangential base points at infinity. 
Its fundamental paths called {\it basic moves} are half-Dehn twists, 
fusing moves and simple moves defined as follows. 

Let $\Delta = (V, E, T)$ be a trivalent graph as above, 
and assume that $\Delta$ is trivalent. 
Then for any rigidification $\tau$ of $\Delta$, 
$\pm E \cup T = \bigcup_{v \in V} {\rm Im}(\tau_{v})$, 
and hence $A_{(\Delta, \tau)}$ is the formal power series ring over ${\mathbb Z}$ 
of $3g + n - 3$ variables $q_{e}$ $(e \in E)$. 
First, the {\it half-Dehn twist} associated with $e$ is defined as the deformation of 
the pointed Riemann surface corresponding to $C_{\Delta}$ by $q_{e} \mapsto - q_{e}$. 
Second, a {\it fusing move (or associative move, A-move)} is defined to be 
different degeneration processes of a $4$-hold Riemann sphere. 
A fusing move changes $(\Delta, e)$ to another trivalent graph $(\Delta', e')$ 
such that $\Delta$, $\Delta'$ become the same graph, 
which we denote by $\Delta''$, if $e, e'$ shrink to a point. 
We denote this move by $\varphi(e, e')$. 
As is done in \cite[Section 3]{I2} and \cite[Theorem 1]{I3}, 
one can construct this move using Theorem 2.2.  
Finally,  {\it simple move (or S-move)} is defined to be different degeneration processes 
of a $1$-hold complex torus. 

Then as the {\it completeness theorem} called in \cite{MS}, 
the following Theorem 2.1 is conjectured in \cite{G} and 
shown in \cite{BK1, BK2, FG, HLS, MS, NS} (especially in \cite[Sections 7 and 8]{NS}  using the notion of quilt-decompositions of Riemann surfaces). 
\vspace{2ex}

\noindent
{\bf Theorem 3.1.} (cf. \cite{BK1, BK2, FG, HLS, MS, NS} and 
\cite[Sections 7 and 8]{NS})   
\begin{it}
The Teichm\"{u}ller groupoid is generated by half-Dehn twists, 
fusing moves and  simple moves with relations induced from 
${\cal M}_{0, \vec{4}}$, ${\cal M}_{0, \vec{5}}$, ${\cal M}_{1, \vec{1}}$ 
and ${\cal M}_{1, \vec{2}}$. 
\end{it}
\vspace{2ex}

\begin{flushleft} 
{\bf 4. Construction of Liouville conformal blocks}  
\end{flushleft}

\noindent
{\it 4.1. Conformal blocks.} 
Fix a real number $c > 1$ called the {\it central charge}, 
and define the {\it Virasoro algebra} ${\rm Vir}_{c}$ with generators 
$L_{n}$ $(n \in {\mathbb Z})$ satisfying the relations 
$$
[L_{n}, L_{m}] = (n-m) L_{n+m} + \frac{c}{12} n (n^{2} - 1) \delta_{n+m, 0}. 
$$
Put ${\mathbb S} = Q/2 + \sqrt{-1} \cdot {\mathbb R}^{+}$, 
and for each $\alpha \in {\mathbb S}$, 
denote by ${\cal V}_{\alpha}$ 
the irreducible highest weight representation of ${\rm Vir}_{c}$ 
with generator $e_{\alpha}$ which is annihilated by $L_{n}$ $(n > 0)$ and 
has the $L_{0}$-eigenvalue $\Delta_{\alpha} = \alpha (Q - \alpha)$, 
where $c = 1 + 6 Q^{2}$. 
Then for any $v \in {\cal V}_{\alpha}$, $L_{n}(v) = 0$ if $n \gg 0$. 
There exists a unique inner product 
$\langle \cdot, \cdot \rangle_{{\cal V}_{\alpha}}$ on ${\cal V}_{\alpha}$ such that 
$$
\left\langle L_{n}(v), w \right\rangle_{{\cal V}_{\alpha}} = 
\left\langle v, L_{-n}(w) \right\rangle_{{\cal V}_{\alpha}}     
\ (v, w \in {\cal V}_{\alpha}), \ \ 
\langle e_{\alpha}, e_{\alpha} \rangle = 1. 
$$

Under $2g - 2 + n > 0$, 
let $C$ be a Riemann surface of genus $g$ with $n$ marked points $P_{1},..., P_{n}$ 
and local coordinates $t_{i}$ $(i = 1,..., n)$ vanishing at $P_{i}$. 
We associate highest weight representations ${\cal V}_{\alpha_{i}}$ of ${\rm Vir}_{c}$ 
to $P_{i}$ $(i = 1,..., n)$, 
and define the action of 
$$
\chi = \left( \sum_{k \in {\mathbb Z}} \chi_{k}^{(i)} t_{i}^{k+1} \partial_{t_{i}} 
\right)_{1 \leq i \leq n} \in \bigoplus_{i=1}^{n} {\mathbb C}((t_{i})) \partial_{t_{i}}
$$ 
on $\bigotimes_{i=1}^{n} {\cal V}_{\alpha_{i}}$ as the following finite sum  
$$
\rho_{\chi} (v_{1} \otimes \cdots \otimes v_{n}) = 
- \sum_{i=1}^{n} v_{1} \otimes \cdots \otimes 
\left( \sum_{k \in {\mathbb Z}} \chi_{k}^{(i)} L_{k}(v_{i}) \right) 
\otimes \cdots \otimes v_{n} \ \left( v_{i} \in {\cal V}_{\alpha_{i}} \right). 
$$
Denote by ${\mathfrak D}_{C}$ the Lie algebra of meromorphic differential operators 
on $C$ which may have poles only at $P_{1},..., P_{n}$. 
Then (invariant) conformal blocks associated with $(C; P_{i}, t_{i})$ are linear maps 
${\cal F}_{C} : \bigotimes_{i=1}^{n} {\cal V}_{\alpha_{i}} \rightarrow {\mathbb C}$ 
satisfying the invariance property: 
$$
{\cal F}_{C}(\rho_{\chi}(v)) = 0 \ \ \left( \chi \in {\mathfrak D}_{C}, \ 
v \in \bigotimes_{i=1}^{n} {\cal V}_{\alpha_{i}} \right),
$$
where $\chi$ is regarded as an element of 
$\bigoplus_{i=1}^{n} {\mathbb C}((t_{i})) \partial_{t_{i}}$ (cf. \cite{FB, T3, T5}).  
If $(g, n) = (0, 3)$, 
then ${\cal F}_{C}$ is uniquely determined by the values 
${\cal F}_{C} \left( e_{\alpha_{1}} \otimes e_{\alpha_{2}} \otimes e_{\alpha_{3}} \right)$. 

Let $C_{i}$ $(i = 1, 2)$ be two Riemann surfaces with $n_{i} + 1$ marked points 
and associated local coordinates, 
and denote by $C_{1} \sharp C_{2}$ the pointed Riemann surfaces obtained by gluing 
$C_{i}$ at their $(n_{i} + 1)$th points via the gluing parameter $q$. 
The gluing of conformal blocks ${\cal F}_{C_{1}}, {\cal F}_{C_{2}}$ by $\beta \in {\mathbb S}$ 
is defined as 
$$
{\cal F}_{C_{1} \sharp C_{2}}^{\beta} (v_{1} \otimes v_{2}) = 
\sum_{l,m} {\cal F}_{C_{1}} (v_{1} \otimes v_{l}) 
\left\langle v_{l}^{*}, q^{L_{0}} v_{m} \right\rangle_{{\cal V}_{\beta}} 
{\cal F}_{C_{2}} (v_{m}^{*} \otimes v_{2}), 
$$
where $\{ v_{l} \}$, $\{ v_{l}^{*} \}$ are dual bases of ${\cal V}_{\beta}$ 
for $\langle \cdot, \cdot \rangle_{{\cal V}_{\beta}}$. 
Then ${\cal F}_{C_{1} \sharp C_{2}}^{\beta} (v_{1} \otimes v_{2})$ is the product of  $q^{\Delta_{\beta}}$ and a formal power series of $q$ with constant term 
${\cal F}_{C_{1}} (v_{1} \otimes e_{\beta}) {\cal F}_{C_{2}} (e_{\beta} \otimes v_{2})$. 
For a Riemann surface $C$ with $n+2$ marked points and 
associated local coordinates, 
the gluing ${\cal F}_{C^{\sharp}}^{\beta}$ of the conformal block ${\cal F}_{C}$ 
can be defined in a similar way, 
where $C^{\sharp}$ denotes the pointed Riemann surface obtained by 
gluing the $(n+1)$th and $(n+2)$th points on $C$. 
Let $\sigma$ be a pants decomposition of a Riemann surface 
of genus $g$ with $n$ marked points and local coordinates, 
and $\beta$ be an ${\mathbb S}$-valued function on the set $E(\sigma)$ of edges 
associated with $\sigma$. 
Then we define the {\it gluing conformal block} ${\cal F}_{\sigma}^{\beta}$ 
as the gluing of conformal blocks on $3$-pointed Riemann spheres, 
and it is represented as a formal power series of deformation parameters of 
the degenerate curve associated with $\sigma$. 

Let $C/S$ be a family of stable curves over ${\mathbb C}$ of genus $g$ 
with $n$ marked points $P_{i}$ and local coordinates $t_{i}$ $(1 \leq i \leq n)$. 
Denote by $\sigma_{i}: S \rightarrow C$ the section corresponding to $P_{i}$. 
Then it is shown in \cite[7.4]{BK2} that in the category of algebraic geometry, 
one can let ${\cal T}_{S}$ act on the sheaf of conformal blocks as follows. 
For a vector field $\theta$ on $S$, 
there exists a lift $\tilde{\theta}$ as a vector field on 
$C - \bigcup_{i=1}^{n} \sigma_{i}(S)$ since it is affine over $S$. 
Take the $i$th vertical component (for the local coordinate $t_{i}$) 
$\tilde{\theta}_{i}^{\rm vert}$ of $\tilde{\theta}$ as 
$$
\tilde{\theta} = \tilde{\theta}_{i}^{\rm vert} + \tilde{\theta}_{i}^{\rm holiz}; \ \ 
\tilde{\theta}^{\rm holiz}(t_{i}) = 0, 
$$
and define the action of $\tilde{\theta}$ on 
$\bigotimes_{i=1}^{n} {\cal V}_{\alpha_{i}} \otimes {\cal O}_{S}$ as 
$$
\theta (f v) = \theta(f) v + f \sum_{i=1}^{n} \rho_{\tilde{\theta}_{i}^{\rm vert}} (v) \ \ 
\left( f \in {\cal O}_{S}, \ v \in \bigotimes_{i=1}^{n} {\cal V}_{\alpha_{i}} \right). 
$$
Then by the definition of conformal blocks, 
this action gives the action of ${\cal T}_{S}$ on the sheaf of conformal blocks on $C/S$. 
We denote $\nabla$ the corresponding connection. 
Then the following result is well known more or less, 
and can be checked using statements and the proof of \cite[Sections 7.4 and 7.8]{BK2}.  
\vspace{2ex}

\noindent
{\bf Proposition 4.1.} 
\begin{it}
\begin{itemize}

\item[\rm (1)]
The connection $\nabla$ is projectively flat. 

\item[\rm (2)] 
The residue of $\nabla$ around the singular locus of $C/S$ is given by the action of $L_{0}$. 

\item[\rm (3)]
The gluing conformal block ${\cal F}_{\sigma}^{\beta}$ gives a (formal) flat section of 
$\nabla$. 

\end{itemize}
\end{it}

\noindent 
{\it Proof.} 
The assertions (1) and (2) are shown in \cite[Proposition 7.4.8]{BK2} and 
\cite[Example 7.4.12 and Corollary 7.8.9]{BK2} respectively. 
We prove (3). 
By \cite[Propositions 7.8.6 and 7.8.7]{BK2} and the proof, 
${\cal F}_{\sigma}^{\beta}$ is the image of a constant section by 
a ${\cal T}_{S}$-equivariant map to the sheaf of conformal blocks over 
$S = {\rm Spec}[[ q_{e} \ (e \in E(\sigma)) ]]$, and hence is a flat section of $\nabla$. 
\ $\square$ 
\vspace{2ex}

\noindent
{\it 4.2. Tempered conformal blocks.} 
We recall results of Teschner \cite{T1, T2, T3} on analytic continuations of 
Liouville conformal blocks on $4$-pointed Riemann spheres. 
We normalize 
$N(\alpha_{1}, \alpha_{2}, \alpha_{3}) = 
{\cal F}_{C} \left( e_{\alpha_{1}} \otimes e_{\alpha_{2}} \otimes e_{\alpha_{3}} \right)$ 
as in \cite[(8.3) and (12.22)]{TV}, 
and $\sigma$, $\sigma'$ be pants decompositions of 
${\mathbb P}^{1}_{\mathbb C} - \{ 0, 1, \infty, x\}$ which are connected by 
a fusing move $x \in (0, 1)$. 
Then it is shown in \cite{T1, T2, T3} that for each $\beta \in {\mathbb S}$, 
the associated conformal block 
$$
{\cal F}_{\sigma}^{\beta} : 
{\cal V}_{\alpha_{1}} \otimes {\cal V}_{\alpha_{2}} \otimes 
{\cal V}_{\alpha_{3}} \otimes {\cal V}_{\alpha_{4}} 
\rightarrow {\mathbb C}
$$
can be analytically continued along $(0, 1)$ to 
a meromorphic form around $x = 1$ which is represented as 
$$
\int_{\mathbb S} d \mu(\beta') \ \Phi_{\beta, \beta'} \ {\cal F}_{\sigma'}^{\beta'}, 
$$
where ${\mathbb S} = Q/2 + \sqrt{-1} \cdot {\mathbb R}^{+}$ 
for a kernel function $\Phi_{\beta, \beta'}$ and a measure $d \mu(\beta')$ 
explicitly given in \cite[5.2]{T2} and \cite[2.1]{T3}. 
Therefore, 
the analytic continuation along $(0, 1)$ gives rise to a canonical isomorphism 
between the Hilbert spaces 
$$
\int_{\mathbb S}^{\oplus} d \beta \ {\mathbb C} {\cal F}_{\sigma}^{\beta} \ \cong \ 
\int_{\mathbb S}^{\oplus} d \beta' \ {\mathbb C} {\cal F}_{\sigma'}^{\beta'} 
$$
obtained as direct integrals. 

The analytic continuation of ${\cal F}_{\sigma}^{\beta}$ 
along a simple move in ${\cal M}_{1, \vec{1}}^{\rm an}$ is given by 
Hadasz-Jask\'{o}lski-Suchanek \cite{HJS}, 
and that along the half-Dehn twist associated with an edge $e$ is the multiplication by 
$\exp \left( \pi \sqrt{-1} \Delta_{\beta(e)} \right)$. 

Using the above results, we will define the space of tempered Liouville conformal blocks. 
Let $C$ be a Riemann surface of genus $g$ with $n$ marked points $P_{i}$ and 
local coordinates $t_{i}$ $(1 \leq i \leq n)$. 
Take a tangential point $p_{\infty}$ at infinity, 
and a path $\pi$ in $\overline{\cal M}_{g, \vec{n}}^{\rm an}$ from $p_{\infty}$ to 
the point $p_{C}$ corresponding to $(C; P_{i}, t_{i})$. 
Denote by $\sigma$ the pants decomposition corresponding to $p_{\infty}$, 
and take an ${\mathbb S}$-valued function $\beta$ on the set $E(\sigma)$ of edges 
associated with $\sigma$. 
For each $v \in \bigotimes_{i=1}^{n} {\cal V}_{\alpha_{i}}$, 
one can see that 
$$
\prod_{e \in E(\sigma)} q_{e}^{-\Delta_{\beta(e)}} \cdot {\cal F}_{\sigma}^{\beta} (v) 
$$
becomes a formal power series of $q_{e}$ $(e \in E(\sigma))$, 
and denote its constant term by $C_{\sigma}^{\beta} (v)$. 
Let ${\cal F}_{C}^{\beta}$ be a conformal block associated with $(C; P_{i}, t_{i})$ 
defined by the condition 
$$
\lim_{p \rightarrow p_{\infty}} \prod_{e \in E(\sigma)} t^{-\Delta_{\beta(e)}} \cdot 
{\rm Trans}_{p}^{p_{C}} \left( {\cal F}_{C}^{\beta} (v) \right) 
= C_{\sigma}^{\beta} (v), 
$$
where $p \in \pi$ approaches to $p_{\infty}$ as $t \downarrow0$, 
and ${\rm Trans}_{p}^{p_{C}}$ denotes the parallel transport by $\nabla$ along $\pi$ 
from $p$ to $p_{C}$. 
Then its analytic continuation as the flat section of $\nabla$ along $\pi$ to $p_{\infty}$ 
has the main part 
$C_{\sigma}^{\beta}(v) \prod_{e \in E(\sigma)} t^{\Delta_{\beta(e)}}$, 
and hence is equal to ${\cal F}_{\sigma}^{\beta}(v)$ as a formal power series 
$\left( \mbox{multiplied by $\prod_{e \in E(\sigma)} q_{e}^{\Delta_{\beta(e)}}$} \right)$. 

We define the space 
${\rm CB}^{\rm temp} \left( \otimes_{i=1}^{n} {\cal V}_{\alpha_{i}}, C \right)$ 
of {\it tempered} conformal blocks associated with $(C; P_{i}, t_{i})$ as the direct integral 
$$
\int_{{\mathbb S}^{3g-3+n}}^{\oplus} \prod_{e \in E(\sigma)} d \beta(e) \ 
\bigotimes {\mathbb C} {\cal F}_{C}^{\beta}
$$
which is isomorphic to the Hilbert space of square-integrable functions on 
$$
\left\{ (\beta(e))_{e \in E(\sigma)} \ | \ \beta(e) \in {\mathbb S} \right\} 
\cong ({\mathbb R}^{+})^{3g-3+n}. 
$$

\noindent
{\bf Theorem 4.2.} 
\begin{it}
\begin{itemize}

\item[\rm (1)]
The Hilbert space 
${\rm CB}^{\rm temp} \left( \otimes_{i=1}^{n} {\cal V}_{\alpha_{i}}, C \right)$ 
is independent of $\pi$ and $p_{\infty}$. 

\item[\rm (2)] 
The Hilbert space 
${\rm CB}^{\rm temp} \left( \otimes_{i=1}^{n} {\cal V}_{\alpha_{i}}, C \right)$ 
satisfies the factorization property in the following sense. 
For Riemann surfaces $C_{i}$ $(i = 1, 2)$ with $n_{i} + 1$ marked points and local coordinates, 
$$
{\rm CB}^{\rm temp} \left( \left( \otimes_{i=1}^{n_{1}} {\cal V}_{\alpha_{i}} \right) 
\otimes \left( \otimes_{j=1}^{n_{2}} {\cal V}_{\alpha_{j}} \right), 
C_{1} \sharp C_{2} \right)
$$
is canonically isomorphic to the direct integral     
$$
\int_{\mathbb S}^{\oplus} d \beta \ 
{\rm CB}^{\rm temp} \left( \left( \otimes_{i=1}^{n_{1}} {\cal V}_{\alpha_{i}} \right) 
\otimes {\cal V}_{\beta}, C_{1} \right) \otimes 
{\rm CB}^{\rm temp} \left( {\cal V}_{\beta} \otimes 
\left( \otimes_{j=1}^{n_{2}} {\cal V}_{\alpha_{j}} \right), C_{2} \right). 
$$
Similarly, for a Riemann surface $C$ with $n + 2$ marked points and local coordinates, 
one has a canonical isomorphism 
$$
{\rm CB}^{\rm temp} \left( \otimes_{i=1}^{n} {\cal V}_{\alpha_{i}}, C^{\sharp} \right) 
\cong 
\int_{\mathbb S}^{\oplus} d \beta \ 
{\rm CB}^{\rm temp} \left( \left( \otimes_{i=1}^{n} {\cal V}_{\alpha_{i}} \right) 
\otimes {\cal V}_{\beta} \otimes {\cal V}_{\beta}, C \right). 
$$

\item[\rm (3)]
By the connection $\nabla$, 
${\rm CB}^{\rm temp} \left( \otimes_{i=1}^{n} {\cal V}_{\alpha_{i}}, C \right)$ 
has a projective action of the Teichm\"{u}ller groupoid for ${\cal M}_{g, \vec{n}}$ such that 
the action of fusing moves and simple moves is given by the action 
in the case when $(g, n) = (0, 4)$ and $(1, 1)$ respectively. 

\end{itemize}
\end{it}

\noindent 
{\it Proof.} 
First, we prove (1). 
Since $\nabla$ is projectively flat, 
${\rm CB}^{\rm temp} = 
{\rm CB}^{\rm temp} \left( \otimes_{i=1}^{n} {\cal V}_{\alpha_{i}}, C \right)$ 
is independent of the homotopy class of $\pi$. 
Then by Theorem 3.1, to prove (1), 
it is enough to show that ${\rm CB}^{\rm temp}$ is independent of 
moving $p_{\infty}$ by fusing moves and simple moves. 
Let $\sigma$ and $\sigma'$ be pants decompositions of Riemann surfaces of genus $g$ 
with $n$ marked points such that $\sigma$, $\sigma'$ are connected by a fusing move $\varphi$. 
Then a gluing conformal block ${\cal F}_{\sigma}^{\beta}$ is represented as 
the gluing ${\cal F}_{C_{1} \sharp C_{2}}^{\beta}$ of ${\cal F}_{C_{1}}^{\beta_{1}}$ and 
${\cal F}_{C_{2}}^{\beta_{2}}$, 
where $C_{1}$ denotes a $4$-pointed Riemann sphere associated with $\varphi$. 
By the above result of Teschner \cite{T1, T2, T3}, 
there exists a form ${\cal F}'_{C_{1}}$ which is the parallel transport of 
${\cal F}_{C_{1}}^{\beta_{1}}$ along the fusing move in ${\cal M}_{0, \vec{4}}^{\rm an}$ 
associated with $\varphi$. 
Then the parallel transport of ${\cal F}_{\sigma}^{\beta}$ along $\varphi$ becomes 
the gluing of ${\cal F}'_{C_{1}}$ and ${\cal F}_{C_{2}}^{\beta_{2}}$ 
by the deformation parameters $u_{i}$ given in Theorem 2.2. 
Therefore, ${\rm CB}^{\rm temp}$ is independent of moving $p_{\infty}$ by fusing moves. 
By the result of Hadasz-Jask\'{o}lski-Suchanek \cite{HJS}, 
the space of tempered conformal blocks for $1$-pointed curves of genus $1$ is 
stable under a simple move. 
Therefore, in a similar way as above, 
one can show that ${\rm CB}^{\rm temp}$ is independent of moving $p_{\infty}$ 
by simple moves.  

Second, we prove (2) in the former case (and the latter case can be shown in a similar way). 
Take pants decompositions of the Riemann surfaces $C_{i}$ $(i = 1, 2)$ 
which give a pants decomposition of $C_{1} \sharp C_{2}$, 
and denote by $p_{\infty}$ the associated tangential point $p_{\infty}$ at infinity. 
Then one can obtain the required isomorphism from the description of 
the space of tempered conformal blocks  associated with $C_{1} \sharp C_{2}$ by $p_{\infty}$. 

The assertion (3) follows from the construction of the space of tempered conformal blocks 
and the proof of (1),  (2). 
\ $\square$

\renewcommand{\refname}{\centerline{\normalsize{\bf References}}}
\bibliographystyle{amsplain}

\end{document}